# Terahertz detection in single wall carbon nanotubes

Kan Fu, Richard Zannoni, Chak Chan, Stephan Adams, John Nicholson, Eric Polizzi and Sigfrid Yngvesson
*Department of Electrical and Computer Engineering, University of Massachusetts, Amherst, MA 01003, USA*

## Abstract

It is reported that terahertz radiation from 0.69 THz to 2.54 THz has been sensitively detected in a device consisting of bundles of metallic carbon nanotubes, quasi-optically coupled through a lithographically fabricated antenna, and a silicon lens. The measured data are consistent with a bolometric process and show promise for operation above 4.2 K.

Single Wall Carbon Nanotubes (SWNTs)[1] have many potential applications in electronics and photonics[2]. Photoconductive detection is apparently very weak[3], but Itkis et al.[4] have reported a sensitive *bolometric* Near Infrared detector based on a Carbon Nanotube (CNT) film. Microwave detection using semi-conducting SWNT (s-SWNT) Schottky barriers[5], s-SWNTs[6], and CNT-FETs[7,8] has been extended to 110 GHz[9]. These results so far indicate no intrinsic frequency limitation due to the CNTs themselves. The question thus arises whether SWNTs could be useful as terahertz detectors. We have proposed a very fast terahertz detector based on the hot electron bolometric (HEB) effect in metallic SWNTs (m-SWNTs)[10]. In this letter we present experimental results on detection of terahertz radiation in m-SWNTs. We interpret our results based on a general bolometric model.

We have previously demonstrated microwave detection in single m-SWNTs[11] that was ascribed to the nonlinearity associated with the "zero-bias anomaly (ZBA)"[12] in the contact resistance (R) at low bias voltage. The microwave response can be predicted from standard microwave detector theory[13]

$$\Delta I = (1/4)*(d^2I/dV^2)*V_{MW}^2 \quad (1) \; ; \; S_I = \Delta I/P_{MW} \; (2); \; S_V = S_I* R \quad (3)$$

Here, $V_{MW}$ is the peak microwave voltage. The factor $d^2I/dV^2$ was calculated from the measured IV-curve. The bias voltage dependence and the magnitude of $\Delta I$ agreed well with that of Eq. (1).

For the present work we have fabricated m-SWNT devices by the Dielectrophoresis (DEP) method[14]. Typically, we apply a 5 MHz voltage of about 5 volts peak to Au contacts made by UV photolithography such as those shown in Figure 1. We used non-conductive sapphire or silicon on sapphire (SOS) substrates. A drop of a suspension of CNTs in isopropyl alcohol is applied in the contact area. The CNTs will then drift to the narrow gap in the contacts and attach to these. The process is halted when the DC resistance is sufficiently low. The result is that a small number of bundles of CNTs will be contacted in parallel. The lower resistance of these devices compared with typical single SWNTs, from 5 kΩ to 50 kΩ, facilitates matching of microwaves or terahertz radiation to the CNTs.

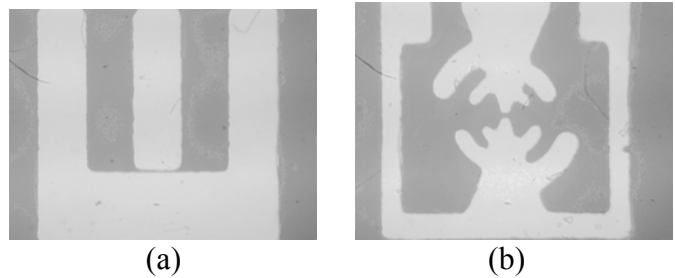

(a)                              (b)

Figure 1. Microwave (a) and Terahertz (b) structures for coupling to the CNTs.

Both structures can be measured in a microwave probe system, a useful diagnostic tool. Figure 1(a) shows a Coplanar Waveguide (CPW) and Figure 1(b) a Log-periodic toothed antenna (LPA). The m-SWNTs are applied across the smallest gaps in these structures, about 4 μm for the CPW and 8 μm for the LPA. The CPW is designed for microwave work, while the LPA is designed primarily for terahertz experiments. Microwave detection with a maximum responsivity close to 1,000 V/W was obtained at room temperature, and found to be essentially frequency independent up to 20 GHz, the highest frequency we have attempted so far.

Each tube is assumed to be modeled by the equivalent circuit introduced and analyzed by P. Burke[15], see Figure 2.

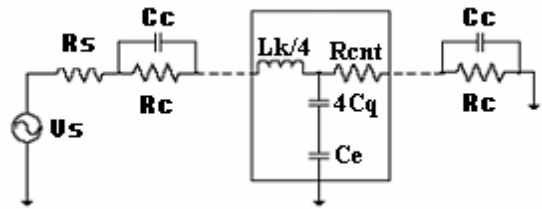

Figure 2. Equivalent circuit model for an m-SWNT[15].

The m-SWNT in the center of Figure 2 is modeled as a transmission line (the unit cells shown are repeated periodically). The contacts are modeled by a resistance parallel with a capacitance. Our microwave characterization work indicates that the total contact capacitance is of the order of 10 fF, in agreement with other recent measurements[16]. The resistance is dominated by the contact resistance and the ZBA effect is still evidenced in the upward curvature of the IV-curves, similar to that found for the single tubes[11]. The propagation velocity on the TL is about 0.01-0.02 times the velocity of light, interpreted as the velocity of a "Tomonaga-Luttinger plasmon" wave. It is predicted that resonances will occur on the TL at terahertz frequencies as its electrical length is a multiple of half wavelengths[15]. Recent theoretical work shows that this model must be modified (faster velocity is predicted) when one considers bundles of m-SWNTs[17].

For the terahertz measurements a device chip with dimensions 6 x 6 mm was inserted in a fixture that allowed quasi-optical coupling to terahertz radiation, as well as bias input and detector output through a coaxial cable and a bias tee. Gold bond wires were used to connect to the contact pads of the LPA. The fixture was mounted in a liquid helium dewar.

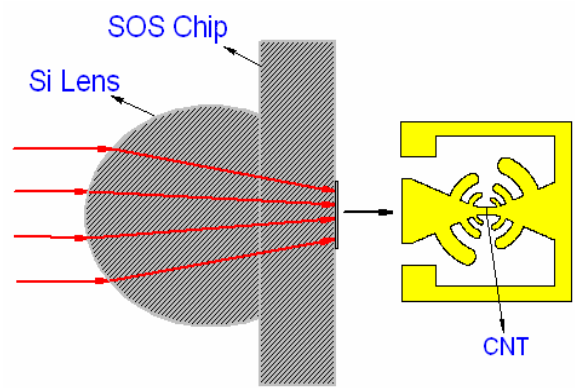

Figure 3. Quasi-optical coupling to the CNTs.

A 4 mm diameter ellipsoidal silicon lens was attached to the substrate for quasi-optical coupling to the antenna[10,18] as shown in Figure 3. The device was biased through a 100 kΩ sensing resistor that configured a Keithley supply as a constant voltage source. A lock-in amplifier was connected across that resistor in a balanced mode in order to record the detected change in current through the device. Terahertz radiation was introduced through the silicon lens from a terahertz gas laser that has a typical output power of 2-5 mW. The laser was modulated at 1 kHz by inserting an acousto-optic modulator after the $CO_2$ pump laser.

Using this configuration we have demonstrated detection at terahertz frequencies in CNT bundles. Five different frequencies were used (wavelength in μm is given in parenthesis): 0.694 THz (432); 1.04 THz (287); 1.395 THz (215); 1.63 THz (184); 2.54 THz (119); We obtained similar responsivities with two devices of quite different resistances: Device A, 430 kΩ, and Device B, initially 7 kΩ, later 20 kΩ, all given at 300 K and low bias voltage. Here we discuss the results for Device B in detail.

This device initially had a room temperature resistance of 7 kΩ ("Device B1") which after about one month changed to 20 kΩ at 300 K ("Device B2"). Many experiments, including several down to 4.2 K, were then performed during which the IV-curves at a given temperature stayed the same. The temperature dependence of the resistance is discussed in detail below. We interpret the step-wise increase in the resistance as being due to a single CNT bundle becoming disconnected.

A summary of all terahertz detections obtained so far is given in Figure 4. The terahertz power was measured outside the window of the dewar, and the response was linear in power. There is a roughly 3 to 4 dB optical loss between the dewar window and the antenna terminals. It is clear that there is a general type of detection process that works for a wide range of terahertz frequencies. The higher resistance device A has more than an order-of-magnitude lower responsivity.

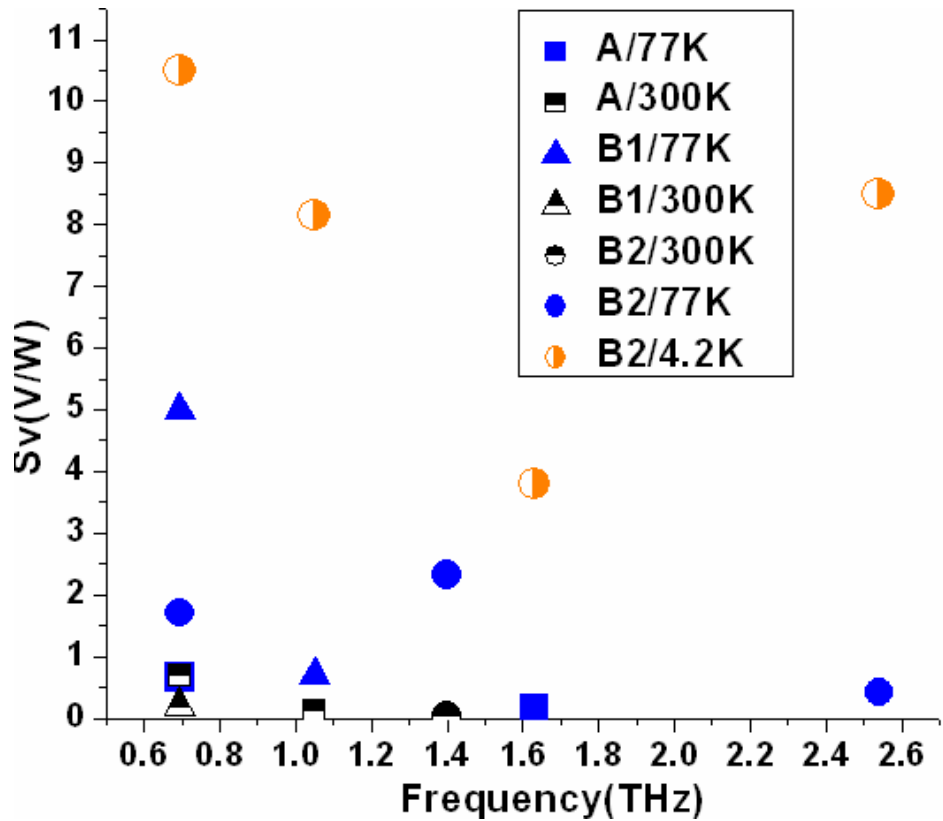

Figure 4. Summary of the responsivities of all terahertz detections obtained, plotted versus frequency.

Our hypothesis is that the detection process at terahertz frequencies is of the bolometric type, similar to that in ref. [4]. A bolometer is a device that has a temperature-dependent resistance R(T) and a heat capacity C. The bolometer is thermally connected through a thermal conductance $G_{th}$ to a heat reservoir at temperature $T_0$. As the bolometer is heated by the terahertz power and biased by the DC current $I_0$, its temperature is increased from $T_0$ to $T_0 + \Delta T$. If we define the factor $b = (1/R)*dR/dT$ then the voltage responsivity of the bolometer will be (neglecting electro-thermal feedback)[19]:

$$S_V = \Delta V / P_{THz} = \frac{I_0 * R * b}{[G_{th} + i\omega C]}(V/W) \qquad (4)$$

The thermal time-constant of the bolometer is determined by $\tau_{th} = C/G_{th}$. We assume slow enough variations in the incident power such that the third term in the denominator in (4) can be neglected.

The Near Infrared bolometer recently demonstrated by Itkis et al.[4], uses a CNT film containing a network of randomly oriented CNTs. This paper showed convincingly that it is essential for achieving a high bolometer responsivity that the CNT film be suspended and not touching the substrate.

We next discuss how a bolometric process can explain our measured data. The equivalent circuit (Figure 2) contains a capacitance parallel to the contact resistance which our microwave probing shows to be at least 5-10 fF, large enough that it effectively shunts the contact resistance at THz frequencies. The equivalent circuit at THz then consists of the transmission line (TL) in Figure 2 only, and this TL has a characteristic impedance of $\sim 10$ k$\Omega$. We estimate that we have about ten metallic SWNTs in parallel. Simulation of the circuit in Figure 3 (for a single m-SWNT) shows that the mismatch loss right at the resonance frequencies may have large peaks, if the damping is weak[20]. For our sample, we expect the m-SWNTs to vary somewhat in length, however, which will tend to wash out the resonances. Moreover, the TL model may be modified by CNT coupling in the bundles[17]. Based on a simple model of ten parallel transmission lines of different length, we find an approximate estimate of the average mismatch loss of 12 dB. Our measured data in Figure 4 are consistent with the simple model, but clearly many further measurements are required for detailed comparison with the theoretical models. Some CNTs may also be quasi-metallic and have a bandgap corresponding to terahertz frequencies[21], which would provide a second efficient mechanism for terahertz absorption in such tubes.

In order to estimate the increase in temperature due to absorption of terahertz radiation we note that Pop et al[22] found a quite large thermal conductance from a single m-SWNT to oxide covered silicon substrates, $g = 0.17$ WK$^{-1}$m$^{-1}$. Similarly, Maune et al.[23] obtained a value of 0.2 WK$^{-1}$m$^{-1}$ for SWNTs on sapphire substrates, as used in this research. We will use the latter value for our estimates below. For ten parallel tubes, 8 μm long, we estimate a total thermal conductance of 1.6 x 10$^{-5}$ W/K at 300 K and 4.1 x 10$^{-6}$ W/K at 77 K. This value is expected to be modified for a bundle of tubes. To further test the model based on Eq. (4) we have measured the resistance of device B2 for a range of temperatures from 300 K to 4.2 K, and calculated dR/dT from this data, see Figure 5 (only a few curves are shown for clarity).

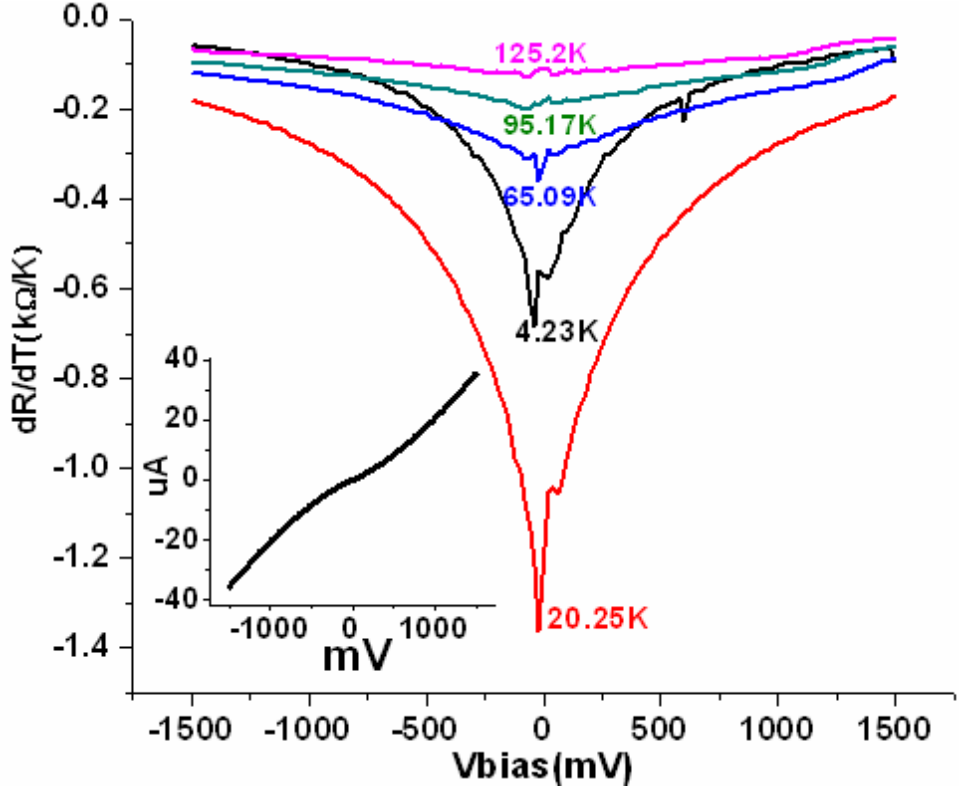

Figure 5. Plots of dR/dT versus bias voltage for several different temperatures. These plots were derived from IV-curves such as that shown in the inset.

We can then predict the bias voltage dependence of the detected voltage response from Eq. (4), while using $G_{th}$ as an adjustable parameter. We find good fits shown below in Figure 6, further supporting the hypothesis that the device detects by a bolometric process. Similar good fits were also obtained at two other laser wavelengths. On the other hand, as is also shown in Figure 6, the measured response does not fit the prediction based on $R*d^2I/dV^2$ (Eqs. (1) – (3)), so the detection processes at terahertz and microwaves are different.

The measured responsivity shown in Figure 6 is based on the power outside the dewar window. If we estimate the total of the mismatch loss and the optical loss to be 16 dB, we then find revised values of $G_{th} = 6*10^{-6}$ W/K (4.2 K); $5*10^{-5}$ W/K (77K). The values for $G_{th}$ at 4.2 K and at 77 K agree within an order-of-magnitude compared with our estimates above. We also measured the temperature-dependence of $S_V$ from 4.2 K to 130 K and based on this found that $G_{th}$ varies as the first power of T from 4.2 K to 50 K, and above that as $T^2$. This is consistent with theoretical expectations such as those in Mingo and Broido[24]. The responsivity decreases by a factor of two at 25 K compared with 4.2 K, indicating a potential for operation above liquid helium temperature.

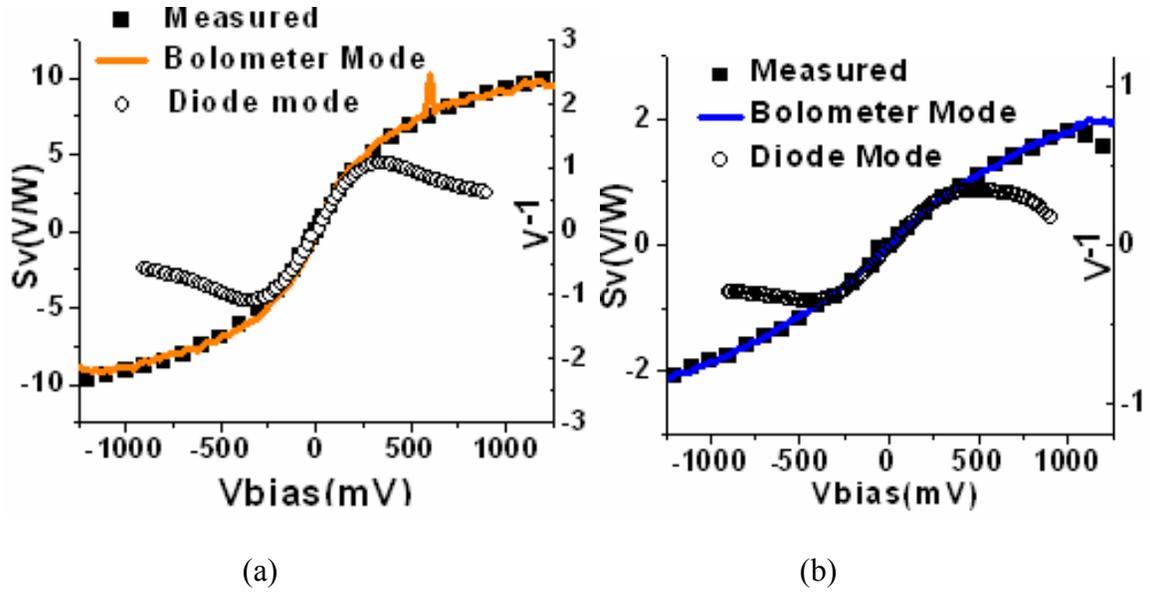

(a)                                                    (b)

Figure 6. Fit of the measured responsivity versus bias voltage to that predicted from Eq.
(4) (bolometer mode) and Eq.'s (1) to (3) (diode mode). All data are for a laser
wavelength of 215 μm (a) At 4.2 K; (b) At 77 K;

To estimate the thermal time-constant we find the heat capacity to be $3.9*10^{-19}$ J/K for a
1 μm long tube at 77 K [25]. For a total of ten tubes, 8 μm long, and the revised values
fitted to the experimental data for $G_{th}$, we obtain $\tau_{th}$ = 1.5 ps at 77 K. A similar value is
estimated at 4.2 K. The measurements do not yet allow us to verify the value for $\tau_{th}$. We
modulate the laser at 1 kHz and find that the detected signal decreases when the
modulation frequency is increased to the range 5 – 15 kHz. This is consistent with the
maximum rate at which our THz gas laser can be modulated, as verified by using a
Schottky diode detector. Future heterodyne detection measurements employing two lasers
will be used to measure the speed of the m-SWNT detector directly. We expect this speed
to be generally consistent with the estimates above.

Based on the work of Itkis et al[4], and Pop et al.[22], we expect the bolometer responsivity
to be much larger (by at least three orders of magnitude) for m-SWNTs suspended in
vacuum across a trench. In that case the thermal conduction will be entirely confined to
the SWNT itself. The thermal time constant will be proportionately longer, and the
responsivity and the thermal time constant may be traded against each other by adjusting
$G_{th}$ as was done in for example[26]. We also note that a simple impedance transformer
could eliminate the mismatch loss over about a 15 % frequency band[10], and thus further
increase the responsivity.

In conclusion, we report detection of terahertz radiation over a wide frequency range in
bundles of metallic CNTs. The experimental data are consistent with a general bolometric
model. While much detailed work remains to clarify and optimize the properties of m-
SWNT terahertz detectors, we have demonstrated the advantage for such work of
employing lens/antenna coupling to the CNTs, as proposed in ref. [10].

This work was supported by NSF grants ECS-0508436 and ECS-0725613.